# Selection, Generalization, and Theories of Cause in Case-Oriented Physics Education Research

Amy D. Robertson, Sarah B. McKagan, and Rachel E. Scherr
Seattle Pacific University, Seattle, WA, 98199

**Abstract**

Case-oriented physics education research – which seeks to refine and develop theory by linking that theory to cases – incorporates distinct practices for selecting data for analysis, generalizing results, and making causal claims. Unanswered questions about these practices may constrain researchers more familiar with the recurrence-oriented research paradigm – which seeks to inform instructional predictions by discerning reproducible, representative patterns and relationships – from participating in or critically engaging with case-oriented research. We use results from interviews with physics education researchers, a synthesis of the literature on research methodologies, and published examples of case-oriented and recurrence-oriented research to answer "hard-hitting questions" that researchers may pose. In doing so, we aim to substantiate our position that both case-oriented and recurrence-oriented PER are rigorous but that the rigor is of a different nature in each paradigm.

## I. Introduction

In a companion paper [1], we describe two distinct paradigms in physics education research (PER)[1] – recurrence-oriented PER and case-oriented PER – distinguished from one another by the assumptions and beliefs about "knowledge, our social world, our ability to know that world, and our reasons for knowing it" that frame and guide each particular orientation toward research, "including what questions to ask, what methods to use, what knowledge claims to strive for, and what defines high-quality work" [2].[2] Recurrence-oriented research seeks reproducible, representative patterns and relationships; human behavior is seen as guided by lawful (albeit probabilistic) relationships. Case-oriented physics education research, in contrast, seeks to refine and develop theory by linking that theory to cases; social action is seen as shaped by the meanings that participants make of their local environments. In delineating these paradigms, we focus on the *premises* and *assumptions* that influence and connect to choices of data and method, rather than on the data or the methods themselves.

Unanswered questions about selection, generalization, and the generation of causal claims in case-oriented research can call its validity into question and constrain productive cross-paradigm dialogue. At a time when our community is increasingly engaging in conversation around issues related to paradigms – organizing invited sessions at national meetings about "Research Paradigms in PER" (Winter 2012 AAPT), "Research directions in PER: Past, present, and future" (2013 PERC), and "Qualitative and Ethnographic Methods in PER" (Winter 2014 AAPT) – we seek to add to the dialogue by answering three of the "hard-hitting" questions that researchers who primarily identify with the recurrence-oriented PER paradigm often ask about case-oriented PER:

1. **Selection:** How do researchers engaging in case-oriented PER select episodes for analysis?

---

[1] In our original paper, we are careful to note that our goal is not to establish that there *only* two research paradigms in PER – *i.e.,* that we have comprehensively filled out a space – but to describe different ways of conceptualizing physics education research that emerged from a reflexive relationship between (i) analysis of interviews that we did with physics education researchers and (ii) literature on research methodologies in the social sciences.

[2] In the same paper, we introduce question-oriented research, an approach that combines the case- and recurrence-oriented research paradigms for the purposes of answering different questions about a single topic, answering different facets of a single question, or creating (and then reconciling) tension between data generated by more than one approach. We do not address question-oriented research in this paper; we refer readers to Ref. 1 for a more detailed description.

2. **Generalization:** On what grounds do researchers engaging in case-oriented PER generalize their results from single cases?

3. **Theories of cause:** On what grounds do researchers engaging in case-oriented PER make causal claims from single cases?

We will argue that case-oriented research and recurrence-oriented research conceive of selection, generalization, and theories of cause in different ways. In particular, whereas recurrence-oriented research makes population-level claims that require researchers to select representative or reproducible data, case-oriented research makes theoretical claims that require researchers to select cases of relevant theory (question 1). When case-oriented research refers to a result that is "general," it often means that a case represents or embodies a theory – that the data is *theoretically significant* – whereas recurrence-oriented research often uses the word "general" to signify that data or a claim is representative of a population of phenomenon (question 2). And case-oriented research often attributes cause to local meanings that are made by participants – inferring cause from process – whereas recurrence-oriented research associates cause with lawful (albeit probabilistic) relationships – inferring cause from controlled experiments (question 3).

Misunderstandings about the differences between the practices of selection, generalization, and causal-claims-generation in case- and recurrence-oriented research can not only translate into missed opportunities for community dialogue, they can also threaten the legitimacy of research, "[a]nd matters of legitimacy have to do with matters of publication, appointment, retention, and promotion" [3]. *Understanding* different approaches has the potential to foster intellectual empathy within the PER community, supporting us in making sense of perspectives and practices that are different than our own; to expand our collective vision for teaching and learning phenomena as we knowledgably take on diverse theoretical and empirical perspectives, "seeing things" through one another's lenses; and to provide us with paradigm-appropriate frameworks for evaluating one another's research.

This manuscript details our answers to the three questions articulated above, arrived at through analysis of interviews with physics education researchers and synthesis of literature on research methodologies. We aim to not only invite researchers more familiar with recurrence-oriented PER to understand selection, generalization, and causal-claims-generation in case-oriented PER, but also to encourage researchers conducting case-oriented research to be transparent about these three practices as they communicate their research results.

More generally, throughout this paper, we take the position that both case-oriented and recurrence-oriented physics education are rigorous. However, the nature of *rigor* – especially with respect to selection, generalization, and the generation of causal claims – is different in each paradigm. Because our community has historically not questioned the rigor of recurrence-oriented PER, this paper primarily seeks to substantiate the legitimacy of case-oriented research. We do so by addressing three case-oriented research practices whose rigor may be called into question, especially as compared to parallel practices in recurrence-oriented research. Throughout the manuscript we regularly connect and contrast these case-oriented research practices with analogous practices in recurrence-oriented research in order to clarify distinctions between the two.

The paper is structured around answering the three questions named above. We begin by discussing the process by which we generated our answers to the three questions in Section II. Sections III, IV, and V lay out our answers to the three questions, in order from selection to generalization to theories of cause. We close with a discussion in Section VI.

## II. Methods

Robertson undertook the current study in her transition from recurrence-oriented to case-oriented PER. She wanted to do more than take up the *methods* of case-oriented PER; she wanted to *experience* case-oriented research, to put on this *lens* as a way of seeing and understanding the learning and teaching of physics. A cursory read of the literature on research methods (*e.g.,* [4,5]) offered her steps and processes, rather than the lens she was seeking, and so, with guidance from Scherr and McKagan, Robertson began to interview physics education researchers, asking questions about why they make the research choices they do. As she conducted and analyzed interviews and familiarized herself with the literature, she realized that specific, unanswered questions about selection, generalization, and the generation of causal claims were undermining her efforts to understand case-oriented PER. Pursuing answers to these questions became a focus of her analysis. In this section, we briefly describe our research process, including our analytic framework.

*Interview sampling.* The interviews Robertson conducted were originally conceived as being in the service of better understanding the research paradigms and approaches described in our companion paper [1]. Early on, we identified two distinct sets of researchers; at the time, we thought of these researchers as conducting either "qualitative" or "quantitative" physics education research, and we chose interview subjects to help us flesh out what each of these meant. As described in our companion paper, what began to emerge as Robertson conducted and analyzed interviews with these researchers was that the features of the research described by her interviewees did not entirely overlap with what is traditionally thought of as *qualitative* or *quantitative* research, especially if conceived in terms of data and methods. Thus, we chose to foreground research paradigms in our analysis, and to call the two paradigms we describe "case-" and "recurrence-oriented," rather than "qualitative" and "quantitative," both to highlight the difference between *paradigms* and *methods,* as well as to foreground what we considered to be a central focus of each paradigm – the priority of recurrence or the importance of cases. As we began to develop our characterizations of the recurrence-oriented and case-oriented PER paradigms, we also noticed that some of the research that our interviewees depicted did not precisely fit either description, and we could think of other examples of prominent physics education research that likewise did not match these descriptions. Subsequently, we conducted additional interviews with individuals whose research we perceived as not fitting into either the case-oriented or recurrence-oriented paradigms. Our characterization of the question-oriented approach (see footnote 2 and original paper [1] for more details) emerged from these latter interviews.

In the end, Robertson conducted eighteen interviews with physics education researchers. Interview participants were chosen not only on the basis of their perceived research interests, but also based on their faculty status, relationship to Robertson, and willingness to be interviewed. Because an original goal of this project was for Robertson to better understand case-oriented research, researchers that we perceived to be conducting this kind of research make up the largest fraction of interview subjects. We gratefully acknowledge the contributions of the following researchers:

Leslie Atkins
*California State University, Chico*

Andrew Boudreaux
*Western Washington University*

Melissa Dancy
*University of Colorado*

Brian Frank
*Middle Tennessee State University*

Ayush Gupta
*University of Maryland*

David Hammer
*Tufts University*

Danielle Harlow
*University of California, Santa Barbara*

Charles Henderson
*Western Michigan University*

Paula Heron
*University of Washington*

Andrew Heckler
*Ohio State University*

Stephen Kanim
*New Mexico State University*

Sarah (Sam) McKagan
*American Association of Physics Teachers*

David Meltzer
*Arizona State University*

Sanjay Rebello
*Kansas State University*

Rosemary Russ
*University of Wisconsin - Madison*

| Mel Sabella | Rachel Scherr | Michael Wittmann |
| --- | --- | --- |
| *Chicago State University* | *Seattle Pacific University* | *University of Maine* |

*Content of interviews.* Each interview lasted between forty-five minutes and one hour and was conducted either in person, by remote video, or on the phone. Interviews were loosely scripted. Major topics included: the kinds of questions each interviewee is interested in answering, the process by which each interviewee tries to answer these questions, the kinds of claims each interviewee seeks to make, what counts as evidence for these claims, and the criteria each uses to evaluate his or her research. Each interview was recorded, content-logged, and summarized. The summaries were sent to individual interviewees and revised on the basis of their feedback.

*Interpretive framework and analysis of interviews.* We identify this manuscript as case-oriented research; the primary goal of our analysis was to understand how the physics education researchers we interviewed make meaning of the practices of selection, generalization, and causal-claims-generation in their own research. The claims we make throughout this paper were generated in a reflexive relationship between (i) Robertson's (originally unarticulated) questions about selection, generalization, and causal-claims-making in case-oriented PER; (ii) subtle but distinctive differences in the way that researchers discussed these three practices in their interviews; and (iii) literature on selection/vision, generalization, and theories of cause in qualitative and quantitative social sciences/educational research more broadly. Upon reflection, we can see that Robertson's own (originally unarticulated) interest in these questions meant that when these topics came up in the natural course of her interviews, she focused on and followed up on them. Likewise, in her content-logging and attempts to understand interviewees' points-of-view, this was a (sometimes unconscious, sometimes more explicit) focus of her interest and attention.

As she analyzed interviews with physics education researchers, Robertson noticed differences in the language that they used to describe the practices of selection, generalization, and the generation of causal claims in their research. For example, some researchers described their research in terms of *theoretical significance,* whereas others did so in terms of *statistical significance.* Some described the answers to 'why' questions in terms of individual volition and meaning-making, whereas others referred to variables that influence or determine particular events or patterns. Concurrent with her analysis, Robertson immersed herself in the methodological literature, and began to see connections between (i) themes in articles and book chapters on selection, generalization, and theories of cause and (ii) statements or research commitments expressed by her interviewees.

Throughout this manuscript, we connect case-oriented PER to qualitative social science research more generally; we do the same for recurrence-oriented PER and quantitative research.[3] Thus, some of the quotes we pull from the literature use the terms "qualitative" and "quantitative." We use quotes from interviews with the researchers named above and examples of published physics or science education research to illustrate current practices of case-oriented and recurrence-oriented research. Published examples were selected because they clearly embody either the case- or recurrence-oriented research paradigm; appear in a journal recognized by our community as a primary site for publishing physics or science education research; and are authored by physics education researchers whose work is recognized as shaping community standards for research.

We use Pollock and Finkelstein's manuscript, "Sustaining educational reforms in introductory physics," [6] as an example of recurrence-oriented PER. These authors build on an earlier study [7] in which they demonstrate that *Tutorials in Introductory Physics* [8], a PER-based curriculum, can be successfully implemented in a new context (*i.e.,* a context other than that in which it was developed). In the study we draw from in this paper [6], the authors explore intra-institutional hand-off of *Tutorials.* They demonstrate that (1) student learning gains [as measured by $<g>$, average normalized gain [9]] are consistently large in courses that implement *Tutorials,* and (2) there is some variation in $<g>$ that may be attributable to curricular choice and/or faculty familiarity with PER.

---

[3] Literature about research methodologies in the social sciences (including educational research) describes qualitative premises with terms such as "interpretive" [13,24,55,67], "constructivist" [4,19,68], or "realist" [11,53] and associates quantitative research with "post-positivism" [4,11,13,19,24,27,55,68].

We identify Berland and Hammer's "Framing for Scientific Argumentation" [10] as an example of case-oriented science education research. This manuscript presents three episodes from a sixth grade science class. The authors focus on the social (*e.g.,* participant roles and appropriate contributions) and epistemic (*e.g.,* what counts as knowledge and how it is generated) expectations constructed by the students and teacher. In the first episode, the "idea sharing discussion," Mr. S (the teacher) nominates students to contribute answers and ideas, often acknowledging or revoicing student ideas without evaluating them. In the second "argumentative" discussion, which takes place later in the unit, students engage with one another's ideas and try "to persuade each other to accept their claims." The "discordant" discussion takes place immediately after the "argumentative" one, when Mr. S "move[s] to resume his role as epistemic and social authority." Unlike in the first two episodes, in the third one, "instabilities" emerge from the competing expectations of students and teacher: some participants' expectations were more consistent with the framing of the idea-sharing discussion, whereas others' expectations were more consistent with the framing of the argumentative discussion. Berland and Hammer connect student and teacher framings in these three episodes to the literature on argumentation and suggest that certain frames are better aligned with productive argumentation practices.

We conducted extensive member checks [4,5,11], offering interview participants and authors of the published examples we use opportunities to provide feedback on drafts of this paper, and we revised the manuscript on the basis of their feedback. Interviewees resonated with our answers to the three questions, although some objected to being specifically labeled as committed to a single paradigm, or to having their research reduced to a single label. To indicate that researchers may simultaneously or sequentially participate in different research paradigms, we focus on *research* commitments, rather than *researcher* commitments. However, in our interviews, several researchers did express strong personal commitments to the premises, values, and assumptions reflected in our characterizations of a specific paradigm, suggesting that some researchers may primarily identify with a single research paradigm at a given time in their career.

## III. Selection: How do researchers engaging in case-oriented PER select episodes for analysis?

Our answer to the question, "How do researchers engaging in case-oriented PER select episodes for analysis?" is that they select episodes by *locating cases of theories in context*. For example, when looking at classroom video, a researcher conducting case-oriented PER may ask him- or herself, "What is this phenomenon a case of?" (rather than, "What are the important patterns or relationships that recur?," as a researcher conducting recurrence-oriented PER might ask). This difference is likely entangled with the different premises and purposes of case-oriented and recurrence-oriented research: recurrence-oriented research that makes *population-level* claims requires representative or reproducible data, whereas case-oriented research that makes *theoretical* claims requires cases of relevant theory. In addition, episode selection in case-oriented PER is likely influenced by assumptions about the complexity and context-dependence of social phenomena; researchers may be especially drawn to interactions that not only illustrate a particular theory or phenomenon but also highlight complexity.

Fleshing out the meaning of our answer advances productive inter-paradigm conversations in a particular way: our experience is that, when asked how they select episodes for analysis, researchers conducting case-oriented PER often say that they pick something that is "interesting" or "striking" from their data, and this may not make sense (and sometimes feels unscientific) to researchers more familiar with the recurrence-oriented PER paradigm. For example, one researcher described the beginning of the case-oriented research process as follows:

> "For me, the process starts when somebody with good professional vision sees something that wows them. The next step is to begin to identify what it is about that that is so impressive. Because I think that a lot of the time as teachers and as researchers, we have sort of a gut reaction, like, 'Look at that! What is that? That is so amazing!' You know, it just gives you chills …"

We will argue that this researcher's sense that an episode is "amazing" is structured in ways that are not immediately obvious – that in fact selection involves extensive (sometimes implicit) theoretical knowledge, and that there are parallels between the ways in which phenomena are selected for analysis in case-oriented and recurrence-oriented physics education research. Below, we describe and illustrate connections between (1) selection practices in case-oriented and recurrence-oriented PER and (2) the specific premises and assumptions associated with each respective paradigm [1]. We also respond to concerns about the subjectivity of case-oriented research and argue that with respect to this construct, the difference between case-oriented and recurrence-oriented research is not the extent to which one or the other is objective but instead is in the researcher's response to the inherent subjectivity of research.

Goodwin [12] defines "professional vision" as "socially organized ways of seeing and understanding events that are answerable to the distinctive interests of a particular social group." The concept of professional vision highlights not only that different social groups have different interests, but also that *specialized ways of seeing are answerable to these differences.* We tie the differences in how researchers conducting case-oriented and recurrence-oriented PER "see and understand events" to the differences in their respective professional visions, which are answerable to distinctive interests (including distinctive aims, premises, and preferences).

## A. Seeing cases of theory

Case-oriented physics education research takes as a premise that the universal properties of teaching and learning are revealed in the details of specific cases. Researchers refine, extend, and refute theories by connecting theory to specific cases, identifying what the case under study is a *case of* [1,13,14]. Researchers conducting case-oriented PER bring their theoretical knowledge to bear (consciously and unconsciously) on the collection and analysis of data: they look for (or tend to see) cases of theory. Wylie [15] (quoted in Freeman, *et al.* [16]) calls this "ladening data with theory." One selection criterion is the extent to which the case is likely to contribute to the development or refinement of theory [17]. Often the articulation of which theory or theories a particular case instantiates is not automatic; the researcher has unconsciously used a certain theoretical lens in her selection, and part of the research process is articulating what that lens is. For example, one researcher we interviewed said:

> "I feel like for me, the process starts when somebody with pretty good researcher or teacher eyes sees something that wows them [and then goes on to] look for other things in the video that maybe seem similar to you, so that you can maybe start to say, "Listen, I don't know what it is, but I feel like these things all go together. I think these are all about the same thing. *What thing are these all about?"* To help you articulate your sense of what matters about the episode."

By "what matters," this researcher means (and agreed that she meant) the episode's theoretical significance, or what theory the episode is a *case of*.

Berland and Hammer's [10] manuscript illustrates the foregrounding of theoretical cases. The authors selected the "argumentative" discussion because it instantiates productive argumentation practices, and the "discordant" discussion stuck out to them because of its relative instability with respect to frames that embed particular social and epistemic expectations. They deliberately searched for a third episode that was more representative with respect to frame stability and argumentation, choosing the "idea sharing" discussion. The authors write:

> "This study grew out of our shared interest in the contrasting dynamics between the argumentative discussion and the subsequent discord: How could we account for the stability of the former and the instability of the latter [with respect to framing]?...We realized [through our analysis…] that we were looking at two idiosyncratic episodes from this class: Although the students and teacher gave the sense that they knew what they were doing in it, the argumentative discussion was unlike anything the first author had observed in this class…, and the discordant episode was unusual in its discord. To get a sense of how things went 'normally,' we examined earlier class discussions and picked two seemingly typical sessions to study through the theoretical lens of framing."

In short, what was 'striking' or 'interesting' to these researchers – what stuck out to them when they watched classroom video – were *cases of* argumentation and framing.

This contrasts with recurrence-oriented research, in which researchers look for (or tend to see) recurrent patterns and relationships in their data. For example, one researcher said:

"We are trying to look at these patterns of results and see whether little variations that seem to be out of the norm in various ways are representative of a significant phenomena that has important ramifications – or has absolutely no relationship to anything significant, and if you do 'em again 99 times, it'll never happen again."

Pollock and Finkelstein [6] notice the "sizable variation of success" in *Tutorials* implementation at CU-Boulder. What was salient to the authors about this variation was its association with the variable of 'course instructor': they observed that students of faculty more familiar with PER outperform students of less familiar faculty. Discussing a table that presents normalized gains on the Force and Motion Conceptual Evaluation [18] for ten different courses, they write, "As seen in Table I, the most significant variations among semesters are associated with the backgrounds of the instructors." Throughout their manuscript, the authors refer to "familiarity"/"background" as a variable that relates to student performance.

### B. Foregrounding complex participant meanings

Case-oriented PER is influenced by premises about social action and the embeddedness of meaning in the details of context. In particular, case-oriented research asserts that (1) social action is shaped by locally constructed meanings[4] and that (2) real phenomena have multiple layers of meaning [1]. Thus, researchers participating in case-oriented PER often foreground (or tend to select) events that (1) highlight the social mediation of meaning and (2) embed nuance and complexity.

Berland and Hammer [10] are explicit about highlighting the social mediation of meaning: they regularly refer to the verbal and non-verbal messages that participants communicate to one another as they reinforce or bid to shift their shared (or unshared) frames. The "discordant" discussion especially stuck out to Berland and Hammer *because of* its instability and the unsharedness of participant framings. Specifically, Mr. S bid to enforce his epistemic and social authority, shifting away from the stable frame in which students were distributing that authority among themselves. Although some students take up his bid, others resist, and this tension between frames manifested as classroom discord that was evident to the authors. Another researcher described her attention to complex classroom events:

"[A]ll of my research [is] organized around trying to appreciate the complexity of what is happening in a fascinating classroom. To really try to put my attention on the incredibly cool stuff that could get missed because it's not part of the assessment and/or the instructor isn't there for it."

---

[4] More specifically, qualitative research in the social sciences (including education research) asserts that people construct locally meaningful interpretations of their environments [4,13,19]; that people take action on the basis of these interpretations (*i.e.,* these interpretations are causal) [13,50]; that culture can organize interactions and promote shared meanings among groups of people that regularly interact [13,69]; and that a primary function of research is to make visible the invisible patterns that organize the participants' experiences [68]. These meaningful interpretations are dynamic and exist only in the context of local interactions, evolving as participants continually make sense of (and shape) their contexts and respond to other participants who are simultaneously making sense of (and shaping) the context [52].

Although we have separated researchers' attention to participant meaning and their attention to complexity, we suspect that the two are actually conflated in the selection process – that researchers likely choose instances of social mediation of meaning that are also complex.

This is in contrast with recurrence-oriented research, which asserts that human behavior is guided by lawful (albeit probabilistic) relationships [13,19] and which privileges recurrence and reproducibility [1,20,21]. Whereas selection in case-oriented research flows out of naturalistic observation of complex social phenomena, selection in recurrence-oriented research is often embedded in experimental design: researchers select variables to test and choose statistical tools or make repeated measurements to ensure representativeness and reproducibility. For example, Pollock and Finkelstein [6] express their interest in understanding whether the positive effects of PER-based curricula (especially *Tutorials in Introductory Physics*) can be successfully reproduced (a) across sites (from development site to new institution) and (b) across faculty within a single institution. Because they want to know if the results are reproducible, they (a) make the same measurements as did the curriculum developers and (b) use the same standard conceptual instruments within their institution.

These recurrence-oriented research premises may also direct researchers' attention toward "clean" data (as opposed to data that is messy or complex). We interpret "clean" to describe experimental data that carefully controls for all but the variables of interest and that meets standards of reproducibility or representativeness, either by repeating the experiment/measurement or by ensuring the sample is random or representative. For example, one researcher said:

> "What I liked about that paper was that it was a very clean study…[I]t was a random handing out of papers. And yet we found a statistically significant difference between how students answered those two questions based on which version they got. The value that I think they have is that they are clean tests of hypothetical models for what's happening with students."

These preferences may help explain why events that are particularly interesting (*e.g.,* complex/messy) to a researcher who primarily identifies with case-oriented PER may be dissatisfying or uninteresting (*e.g.,* disorderly) to a researcher who identifies primarily with recurrence-oriented PER.

### C. Responding to concerns about subjectivity: Making bias visible

Some researchers who primarily identify with the recurrence-oriented PER paradigm may be less concerned about *how* episode selection takes place in case-oriented PER – *what* influences researchers' vision – and instead may be more concerned that episode selection by its very nature – selecting cases to illustrate or refine theory – is subjective. Our sense is that this concern is tied to a view (common in recurrence-oriented research) of subjectivity as something to be guarded against and to associated standards and practices for minimizing the subjectivity of research results (which case-oriented research often does not meet). In case-oriented research, however, subjectivity is understood to be inevitable, and associated standards and practices focus on explicit acknowledgment of various sources of bias or influence. In what follows, we argue that the difference between selection practices in case-oriented and recurrence-oriented PER is not the degree of subjectivity but the researcher's response to it.

Both selection and invention occur at multiple stages of the case-oriented and recurrence-oriented research processes described by our interviewees. For example, in case-oriented research that seeks to construct narratives of particular classroom events, researchers make selections when they: choose when and where to record video (which entails selecting relevant populations or content); capture video (which involves pointing the camera in a particular direction [22]); select an episode (which involves choosing a portion of the video corpus to analyze in detail); and formulate claims (which involves highlighting particular parts of the selected episode as evidence). These same researchers participate in invention as they build connections between case and theory (in order to articulate and refine what a particular episode is a *case of*), as well as when they categorize and interpret observations to formulate claims. In recurrence-oriented research that aims to characterize student conceptual understanding and develop research-based curriculum, researchers make selections as they: choose relevant content on the basis of which to write a question (which involves selecting a topic and, in some cases, deciding what are the learning goals for that topic); formulate conceptual questions (which entails choosing some feature of the selected content on which to focus); administer these conceptual questions (which involves selecting relevant populations and settings); formulate interpretive categories (which involves selective attention to the data); and categorize student responses. Invention takes place when, for example, these researchers design hypothetical scenarios for conceptual tasks and when they formulate interpretive categories (and in doing so, discover meaning in student responses).

The literature on research methodologies responds to the prevalence and necessity of invention and creativity by addressing the role of the researcher, arguing that it is not his or her task to ensure that another would discover the same meaning in the same data [20]. Rather, it is the researcher's task to make explicit why he made the choices he did and why those choices were reasonable ones to make. Marton [23], discussing phenomenographic research, says:

> "… someone usually asks: Would another researcher working independently arrive at the same set of categories if he or she were studying the same data? On the surface, this appears to be a reasonable question. After all, research results are supposed to be replicable. However, two issues are buried in the question. One concerns the process of discovery: Would other researchers find the same conceptions or categories if they were doing the study for the first time? (Analogously, we might ask, Would two botanists discover the same plants and species if they independently explored the same island?) The other issue concerns whether a conception or category can be found or recognized by others once it has been described to them by the original researcher. The point I want to make is that replicability in the second sense is reasonable to expect, but in the first sense it is not. The original finding of the categories of description is a form of discovery, and discoveries do not have to be replicable. On the other hand, once the categories have been found, it must be possible to reach a high degree of intersubjective agreement concerning their presence or absence if other researchers are to be able to use them. Structurally, the distinction I draw here is similar to that between inventing an experiment and carrying it out. Nobody would require different researchers independently to invent the same experiment. Once it has been invented, however, it should be carried out with similar results even in different places by different researchers."

We suspect that most physics education researchers would acknowledge that their selection and analysis practices involve the kind of subjectivity we describe above. However, researchers participating in the two different paradigms often respond very differently to this research reality [24]. Case-oriented research argues that it is impossible to eliminate the influence of the researcher on the research and so researchers seek to articulate the theories and perspectives that affect (or bias) their selection and analysis [16,25]:

> "The researcher brings ways of thinking about learning, about physics, about learning physics, and about how people do (or should) behave and think. A good qualitative researcher acknowledges the many subjectivities or tacit theories that, even though he may not be aware of, guide his actions (where to point the video camera, what he notices going on in class, what type of data he decides to collect)." [5]

Recurrence-oriented research, in contrast, seeks to eliminate "selection bias" [26] by privileging those phenomena that recur over and over, independent of observer and context:

> "Even if observations are never theory-neutral, many of them have stubbornly re-occurred whatever the researcher's predilections…So, even if there are no 'facts' we can independently know to be certain, there are still many propositions with such a high-degree of facticity that they can be confidently treated as though they were true." [21]

These very different approaches entail (and are entailed by) quite different perspectives about what constitutes rigorous work:

"If you believe, for instance, that good scientists should be objective in the sense of producing knowledge that is epistemologically independent of their personal values and sociopolitical beliefs, then you are likely to privilege as rigorous those methods that demonstrate agreement (replication or reproducibility) among independent observers. If, instead, you believe knowledge is unavoidably shaped by the preconceptions of the knowers (and that independent agreement may simply be a manifestation of a shared bias among the members of a research community), then you are likely to privilege as rigorous those methods that illuminate the nature of the bias and the social, cultural, and political factors that shaped it." [24]

These differences in perspective are connected to the different assumptions about social phenomena/human behavior taken by case-oriented and recurrence-oriented research. In particular, if one assumes, as does case-oriented research, that the important causal influences on human action and behavior are the context-sensitive meanings made by participants, it makes sense that one would respond to the subjective influence of the researcher by seeking to make visible the "historical, cultural, institutional, and immediate situational" [24] influences on the meanings one is making of one's own research. In contrast, the recurrence-oriented research assumption that there are predictable, causal relationships that govern human behavior suggests that these relationships can exist completely apart from our understanding or study of them. In this case, probability and statistics or repeated observations are seen as providing a means to assess or minimize the likelihood that the phenomenon has been mischaracterized or misunderstood [3,27].

Concerns about the subjectivity of case-oriented research are likely tied not only to bias in the selection of cases but also to questions of generalizability – concerns that a single case is insufficient to make generalizable claims, or that claims made on the basis of a single case are naturally subjective. These concerns may be grounded in the perspective that "generalities" are those patterns and relationships that recur across cases and that research results are trustworthy (*i.e.,* represent real results that can inform predictions) when they recur over and over. From such a perspective, the analysis of a single case may be seen as subjective in the sense of generating results that may be biased or random. In Section IV, we will argue that case-oriented research takes a different perspective on generalizability and on what is trustworthy and useful, one that is intimately linked to the study of cases.

## IV. Generalization: On what grounds do researchers engaging in case-oriented PER generalize their results from single cases?

In response to the question, "On what grounds do researchers engaging in case-oriented PER generalize their results from single cases?," we answer that case-oriented research assumes a perspective in which the universal is manifested in the particular details of specific cases, rather than (or in addition to) in the recurring patterns that emerge across cases [13,14]. With this perspective, it is not only possible but *necessary* that researchers look to cases to find universals (*i.e.,* to make generalizable claims). This perspective is clarified in the following excerpt from the literature on research methodologies:

"Mainstream positivist [quantitative] research on teaching searches for general characteristics of the analytically generalized teacher. From an interpretive [qualitative] point of view, however, effective teaching is seen not as a set of generalized attributes of a teacher or of students. Rather, effective teaching is seen as occurring in the particular and concrete circumstances of the practice of a specific teacher with a specific set of students 'this year,' 'this day,' and 'this moment'…This is not to say that interpretive [qualitative] research is not interested in the discovery of universals, but it takes a different route to their discovery, given the assumptions about the state of nature in social life that interpretive researchers make. The search is not for *abstract universals* arrived at by statistical generalization from a sample to a population, but for *concrete universals,* arrived at by studying a specific case in great detail and then comparing it with other cases studied in great detail. The assumption is that when we see a particular instance of a teacher teaching, some aspects of what occurs are absolutely generic, that is, they apply cross-culturally and across human history to all teaching situations...These [universal] properties are manifested in the concrete, not in the abstract." [13]

Erickson's explanation that "some aspects of what occurs [in a particular instance of a teacher teaching] are absolutely generic" corresponds to the idea that, in a given moment, the teacher instantiates a theory (or several) – that some of her actions are a *case of* a more general theory. Thus, case-oriented PER uses the word "general" to signify representative of or embodying a theory. In this paradigm, generalizations are made by connecting case to theory. In contrast, in recurrence-oriented PER, the word "general" typically means representative of a population or phenomenon; generalizations are made by identifying patterns and relationships that emerge across cases.

The question of generalization is particularly significant to understanding case-oriented research: for many researchers, generalization is the *purpose* of research. Unarticulated paradigmatic differences in the meaning of generalization or in assumptions about where universals are located, how to locate such universals, and how research generalizations might serve the broader community may promote misinterpretations of case-oriented research as ungeneralizable or without purpose.

The case-oriented research perspective on generalization is intimately connected to premises about the context-dependence of social action. In addition, the original question is logically connected to a series of follow-up questions, including how one discerns the universal from the particular (by connecting case to theory) and how such universals are useful to other contexts (by broadening awareness). In the following sections, we will discuss these perspectives and questions, connect the use of cases to the goals of case-oriented research, and illustrate that the case-oriented research perspective on generalization fleshes out in specific kinds of claims.

### A. Single cases retain necessary complexity

For the researcher engaged in case-oriented PER, knowledge of theory is inseparably linked to understanding of context-rich cases. Case-oriented research asserts that people construct *locally* meaningful interpretations of their environments and that people take action on the basis of these interpretations (*i.e.,* the interpretations are causal) [4,13,19,28]. Taken out of context, these meanings *lose their meaning,* since they are so intricately entangled with the experiences and environments in which they are made. Thus, in case-oriented research, the reality at the core of any analysis is the meanings that participants make of their experiences and environments.

In our interviews, researchers who primarily engage in case-oriented PER expressed skepticism toward analyses that do not include details of specific cases. Such analyses are thought to oversimplify or to abstract so far as to lose their utility, since reality embeds so much complexity (made apparent in the context-rich details of specific cases). For example, one researcher says:

> "We started watching video of what was going on during the tutorials, and on video it was totally different than what I thought. There was just so much going on in there that I had been unaware of. My former thing about pre- and post-testing was just missing so much of the amazing stuff that was *really* happening in the tutorial. I feel like ever since, all of my research has been organized around trying to appreciate the complexity of what is happening in a fascinating classroom."

This researcher's sense that the recurrence-oriented research that she conducted in her early career missed what was "really happening" communicates the importance of providing context-rich details to her own research and her mistrust in research that simplifies classroom events.

Recurrence-oriented research, on the other hand, embodies different premises about what is real [1]. In this paradigm, what counts as real are patterns and relationships that recur across contexts. One researcher said,

> "It strikes me that from some of the pure qualitative research that I see, it looks like you see an incident and then you try to understand what was going on in that incident. And then making some leap to say that because this happened in this incident, we should be aware of such and such, or pay attention to this or that…It hasn't established…anything more than that…[What is different about my approach is that I] try to ensure that what is coming out is not *only* an artifact of the question […by asking] a variety of different questions…[I]f you get similar responses to different questions, […you know that] you've got something that is more than just a reaction to a particular question. Like there's something there. It may not be completely robust and coherent, but it's gonna come up under a variety of circumstances."

Such statements imply that case-oriented research does not reflect how things "really are" because the recurrence or reproducibility of the results has not been established. This notion that rigor is comprised of establishing recurrence and context-independence is corroborated by a statement made by a second researcher, who said:

"In general, I say, 'That's a very interesting result. Now do it again and see what happens.' And if you get it a third time and if it's similar to what you observed the first two times, then you can begin to believe that you're onto something."

Researchers engaged in case-oriented PER do not necessarily *lack* trust in recurring phenomena, if those phenomena are seen and interpreted in context. In fact, these researchers expressed their hesitancy to report something that they have seen only once; they will rarely share an episode that has not been corroborated or refined by another episode. Rather, they will likely choose the best case of a relatively common thing they have seen – the one that has the greatest potential to expand or refine the theory it illustrates. One researcher engaged in case-oriented PER says, "Usually I've seen it as a recurring pattern, at least anecdotally, and then I'm trying to say more about that case in particular."

### B. Single cases connect to theory

If one accepts the premise that that the general can be manifested in the particular, the next logical question might be *how* one discerns the general from the particular. In qualitative research more broadly, and in case-oriented PER, researchers see their task as

"uncover[ing] the different layers of universality and particularity that are confronted in the specific case at hand – what is broadly universal, what generalizes to other similar situations, what is unique to the given instance." [13]

In other words, researchers first discern what the case under study is a case of, and then they build connections between case and theory (*i.e.,* the universal).[5] We sense that this is what one of our interviewees means when she says:

"I feel like for me, the process starts when somebody with pretty good researcher or teacher eyes sees something that wows them [and then goes on to] look for other things in the video that maybe seem

---

[5] While not the focus of many of our interviews, qualitative or case-oriented research may also generate theory via across- or within-case connections [13]. Although a sort of pattern-seeking, this practice differs from searching for patterns in aggregate data, both in its assumptions and in its approach. Researchers who make within- or across-case connections maintain the assumption that the universal is manifest in the concrete. They seek to flesh out each case in great detail, developing a model or generating claims about what is going on within each specific context. Comparing and contrasting cases serves to deepen the researcher's understandings of the cases themselves – individually and collectively – often leading to iterative refinement of claims about what is going on within and across cases. For example, Schofield [20] makes inferences about the generality of particular desegregation phenomena by making connections within and across cases. She writes that students in the specific school she studies "develop[ed] a color-blind perspective" and a "taboo against the mention of race." She also noticed "asymmetrical concerns" among black and white students: the white children "perceived blacks as something of a threat to their physical selves," whereas the black students "perceived the white as a threat to their social selves." She calls the former "atypical" of desegregated schools, since the conditions for the development of a taboo "seemed to be linked to the special characteristics of the school" (illustrating within-case connections). The latter, on the other hand, she calls likely "more widespread," since the asymmetrical concerns seem "linked to the black and white students' situation in the larger society and to powerful historical and economic forces, not to special aspects of the school" (across-cases-of-desegregation connection). She goes on to clarify that such asymmetrical concerns may look different in different social/school contexts, consistent with the case-oriented paradigm's emphasis on locally constructed meanings.

similar to you, so that you can maybe start to say, "Listen, I don't know what it is, but I feel like these things all go together. I think these are all about the same thing. *What thing are these all about?"* To help you articulate your sense of what matters about the episode." (quoted above)

These connections between case and theory reveal ways in which the universal (theory) is manifested in the concrete (case), providing opportunities for clarification of the theory itself and, in some cases, pointing researchers to new theoretical territory (*e.g.,* when the theories that are brought to bear cannot account for the details of the case) [29,30]. Without these case-theory connections, selection and analysis of episodes does not produce generalizations; there are a nearly limitless number of possible – even true – claims that can be made about any given episode, but a limited number have theoretical significance, or "matter," in the words of our interviewee.

Eisenhart [17] offers an illustration not only of the kind of theory that case-oriented research seeks to discover but also of how connections (i) between cases or (ii) between case and existing theory clarify the theory itself. She excerpts Becker's [31] summary of in-depth studies of men's prison culture and efforts to generalize these results to a women's prison. In her description, studies of men's prisons demonstrated that inmates develop an elaborate prison culture, including a market for scarce material goods and personal services and a code of conduct emphasizing the protection of information. Researchers attributed these inventions to the deprivations of prison life. Other researchers, with this theory in mind, studied a women's prison, but found no comparable code of conduct and a very different structure of social relationships. Rather than invalidating the original theory that the deprivation of prison life led to the creation of a prison culture, the new study extended the theory, recognizing that prison deprives women of different things than men (*e.g.,* familial protection rather than autonomy) and that "[t]heir culture responds to that difference."

Knowledge of how researchers conducting case studies discern the universal from the particulars of a given case gives us insight as we evaluate and seek to understand research in our own field. For example, this knowledge supports us in making explicit the process by which Berland and Hammer [10] build connections between case and theory. These authors articulate what their three episodes are *cases of,* saying:

> "The empirical case we present shows multiple stabilities in the students' and teacher's understandings of what is taking place during argumentative and more traditional class discussions, with dynamics at the levels both of individuals and of the class as a whole. The theoretical case we present is that these phenomena of student, teacher, and class dynamics connects to prior work on *frames* and *framing* [citations from framing literature]..."

Throughout their analysis, they connect the dynamics of their cases to the literature on framing, offering their readers a particular theoretical lens through which to view the discussion. They also point out the ways in which the students instantiate (or do not instantiate) scientific argumentation practices in each discussion. They use their cases to inform theory by showing how specific framings do or do not align with productive argumentation practices. Consistent with case-oriented research premises, this connection is forged through *context*: the authors point to literature that suggests that certain contexts – or certain ways that students "experience their context" – support and call for scientific argumentation practices, and they show that the meanings Mr. S' students are making of their classroom discussions support and/or call for such practices. Berland and Hammer also connect case to case, drawing parallels and distinctions between the framings, dynamics, and argumentation practices in the three episodes.

This means of discovering universals contrasts with that in quantitative social sciences research and in recurrence-oriented PER, in which the task of the researcher is to uncover patterns and relationships that are representative (and thus predictive) of some phenomenon or population. In practice, this is often difficult, since it requires random sampling (*i.e.,* sampling that ensures that each member of a population has an equal chance of being selected) and, in some cases, random assignment (*i.e.,* each member of a sample has an equal chance of being assigned to the control and experimental groups) [32]. When these requirements cannot be met, researchers look for patterns and relationships

> "us[ing] a different generalization model, one that emphasizes how consistently a causal relationship reproduces across multiple sources of heterogeneity...The operative question is: Can the same causal relationship be observed across different laboratories, time periods, regions of the country, and ways of operationalizing the cause and effect?" [21]

The model that Cook describes maps onto several of Pollock and Finkelstein's [6] claims. For example, as discussed earlier, the authors seek to discern whether the effectiveness of *Tutorials* is reproducible across institutions and intra-institutional implementations. This is particularly relevant to claims of representativeness: neither the original implementation of *Tutorials* (by the curriculum developers) nor the original implementation at CU-Boulder (the authors' institution) represent typical implementation conditions. To address questions about representativeness, the authors reproduced the measurements across and within institutions. The authors found that the gains posted (1) by CU-Boulder and (2) by non-PER faculty at CU-Boulder were consistently large, which suggests that the *Tutorials* may be *generally* effective.

### C. Single cases illustrate theories and broaden awareness

Even if one understands *that* case-oriented research subscribes to a different model for generalization and understands *how* researchers go about discerning the universal from the particular, one may still be left with the question of how such research is *useful* if it is derived from a single case or context. Our answer to this question is that just as case-oriented and recurrence-oriented research (1) take different perspectives on where universals manifest and (2) take different stances toward cause in the social world/human behavior, researchers participating in these two different paradigms also think of the usefulness of research results differently. Researchers conducting case-oriented PER want their research results to be useful in an *awareness* sense, rather than in a *representative* sense. For example, one researcher whom we interviewed stated his goal as "expanding the perspective" of his readers:

> "My goal is to say at the end of the day that I've expanded the perspective that a reader of my work might have. That at the beginning they wouldn't have thought that this [event] was something of interest, then notice this set of things that happened, the complexity of it, the richness of it, and notice how much we could be paying attention to or are paying attention to when we naturally interact with the world."

In particular, case-oriented research broadens readers' awareness by illustrating what a theory *looks like* in context, which may heighten readers' theoretical vision in their own contexts, or by refining theory, adding to readers' existing theoretical understanding. Because researchers who strongly identify with the case-oriented PER paradigm believe that it is impossible to separate one's theoretical commitments from one's vision of and for events, they believe that providing readers with new theoretical lenses necessarily shapes what they see and how they see it.

The research literature affirms this purpose of case-oriented research, calling it particularly appropriate for enhancing readers' awareness of (and in) situations similar to the one studied [33,34]. The situation need only be similar in that it instantiates the same theory; it may be very different in other ways. Using Eisenhart's illustration above, the theory that "prisoners develop a culture that solves the problems created by the deprivations of prison life" can explain the cultures of both the men's and women's prisons, even though the specific cultures developed in men's and women's prisons look very different in practice. Erickson [35] says that for qualitative research,

> "judgments of external validity [generalizability] lie in the eye (and experience) of the reader. If you as a reader recognize in my descriptions processes you find also at work in settings you know, then you are determining that what I am saying below 'generalizes' beyond the cases I am reporting."

Wehlage [36] calls this "generalization by analogy":

> "The implication of this view is that generalization is more like thinking by analogy than discovering law-like empirical relationships…[O]ne is to use the data from the case study as an example of the kind of thing that happens in situations like that…Despite some difficulties with this concept, there is strong appeal in the notion that field studies can provide us with a broader range of (surrogate) experience than we could otherwise have, and that the generalized insights that we can take from this experience will help us to act more intelligently in future contexts."

Berland and Hammer's [10] manuscript reflects the position that theory development is especially suited for broadening the reader's perspective and sharpening awareness. In particular, the authors highlight the "productive resources for argumentation" that students in Mr. S' class bring to bear in the "argumentative" discussion. They suggest that educators recognize and "tap into" these resources by fostering contexts that "students recognize as argumentative." By illustrating what such resources can look like, Berland and Hammer enhance readers' vision for their own students' resources for argumentation, and by proposing a theoretical connection between framing and argumentation, they broaden readers' perspective of how they might foster argumentation (through certain framings) or for what may be constraining the engagement of their students in argumentation. Berland and Hammer do not prescribe a specific structure for contexts that promote argumentation; in fact, they warn against focusing on "steps or components of argumentation." Instead, they recommend that instructors support students in achieving results that are meaningful to them.

Striving for usefulness through awareness is more appropriate to case-oriented research than is striving for representativeness, given the case-oriented research premise that social action is guided by *locally* constructed meanings. What the researcher hopes the reader will become aware of is the universal – for example, that people coordinate multiple modalities to communicate understanding, not that *specific* participants will coordinate *specific* modalities to communicate a *specific* understanding. In other words, the aim is not to establish that the universal would manifest in a predictable way in another setting. One interviewee said:

> "…if [another researcher is] interpreting my work to be about [predictability], then no wonder [they] *ought* to be disappointed, because that's not it. I think of my research as being much more likely to illustrate the mechanism by which some learning process occurs. It would be about a small number of situations, and it would hopefully make a convincing case that that can be how things work, but not that it is how things would work next time."

Recurrence-oriented research, on the other hand, aims to offer readers results that are useful because they are *representative* and can thus inform predictions about a particular population:

> "You'd like to be able to make generalizations about *some* population that is reproducible. It doesn't have to be average, it could be the above-average students. You could say, 'We have studied the above-average students, but we believe that these results are representative of a definable group, more than in the spring semester of my Physics 102 course at X university.' [I]t's of no interest to anyone else unless you can show how it relates to their situation. So science works on generalizable results."

Thus, if one wants to know whether a recurrence-oriented study is useful for one's own context, one must first figure out whether one's sample is representative of the same population as the sample in the study. If this is the case, the same parameters and relationships should apply. This concern for representativeness as a critical element of cross-contextual generalizability to other contexts (*i.e.,* external validity) is echoed by the literature on (quantitative) experimental validity [37-40].

Pollock and Finkelstein's [6] paper reflects the 'representative' sense of the usefulness of research. They offer extensive demographic information and implementation specifics – information that others could use to compare their sample to the authors'. They choose <g> as their measure of student achievement, a construct that has been shown to be independent of students' pre-test scores [9] (thus normalizing their pre-instructional state). They anticipate that some readers will call into question the *representativeness* of their claim that the *Tutorials* are effective and that these readers may instead think that the effectiveness of the *Tutorials* is attributable to well-prepared faculty. They acknowledge the limitation of their study in explicitly addressing this concern, saying, "Of course, the most compelling study would be to control for faculty member and vary the curricula. However, we have not had opportunity to [do so]." However, they point to the consistent positive effects of the curriculum *across* instructors, marshaling this recurrence as evidence that "the materials themselves…are important."

### D. Single cases address certain types of research goals and questions

Part of what makes case study research appropriate for the discovery of specific kinds of generalizations is its alignment with the goals of case-oriented research, as reported by our interview participants (see [1]). For example, case-oriented physics education research seeks to broaden audience perspective by illustrating, building, and refining theories. Researchers clarify participants' points of view, reveal and challenge implicit assumptions, demonstrate possibility, develop mechanisms that explain certain teaching and learning phenomena, and coordinate multiple modalities to better understand thinking and learning. Recurrence-oriented physics education research, on the other hand, seeks to help readers plan and predict instruction by identifying recurring teaching and learning phenomena – such as conceptual difficulties that students may encounter when learning concept $x$ – and instructional causes and effects – such as variables that influence learning gains and misconception-like patterns in student responses. We can logically connect each of these aims to a researcher's choice to study single cases or representative populations. For example, if one seeks to reveal and challenge implicit assumptions, one need only deeply study a *single* (detailed) instance that contradicts a standard assumption. (The goal in doing so is to refine the assumption, not simply to demonstrate the contradiction.) Likewise, to demonstrate possibility – *e.g.,* to show that a type of interaction is possible or that a type of learning *can* happen in a science classroom – only requires a single instance. On the other hand, large $N$ is necessary if one wants to make claims about instructional effectiveness for a *representative* student or population. Similarly, if one seeks to identify prevalent student ideas or get a broad sense of the set of ideas that introductory physics students may have about a particular concept, one must ask questions of large numbers of students.

Tables I and II summarize the claims made by eight additional examples of case- and recurrence-oriented PER. In each case, we provide the evidence that authors cite for their claim(s) and the significance of the analysis as a whole.

**Table I.** *Claims and evidence from examples of case-oriented physics education research*

| Author and title of paper | Major claim(s) | Evidence for claim(s) | Significance: Informs theory about? |
|---|---|---|---|
| Harlow, "Structures and Improvisation for Inquiry-Based Science Instruction: A Teacher's Adaptation of a Model of Magnetism Activity" [41] | (1) Ms. Carter (elementary school science teacher) revises a research-based "Models of Magnetism" PET [42] activity to test the models for a magnetized nail that her students propose (rather than the "charge separation" model that the curriculum anticipates). The "*differences* [in the enacted curriculum] at the event level were necessary to preserve the *similarities* at the scientific practice level." (2) Ms. Carter drew on knowledge of "the nature of science, children's learning, and science content" as she transformed the activity. | (1) Ms. Carter asks her students to conduct experiments that (i) challenge their proposed models for a magnetized nail and that (ii) are different from the experiments proposed by the "Models of Magnetism" unit. (2) Ms. Carter suggested her students test their models, deviated from the experiment planned by the curriculum, and proposed a model that challenged her students' models. | (1) Role of adaptive instruction in inquiry-based learning (2) Pedagogical content knowledge entailed by scientific inquiry in the classroom |
| Gupta, Elby, and Conlin, "How substance-based | "Despite the seeming unproductiveness of a substance ontology of force, we argue that thinking of gravity as | (1) Lynn's and Daniel's use of substance-like reasoning about gravity fed into the correct | Role of misontologies in learning |
| ontologies for gravity can be productive: A case study" [43] | 'stufflike' contributed to learners' conceptual process in learning about gravity, forces, and motion, progress that would likely have been less transformative for the participants had the instructors steered learners away from this 'misontology.'" | "Newtonian compensation" argument for why more and less massive objects fall at the same rate. (2) Lynn cites "figuring it [the answer to this question] out myself" as a transformative experience. | |
| Lising and Elby, "The impact of epistemology on learning: A case study from introductory physics" [44] | Jan (a student enrolled in an introductory physics course at the University of Maryland) experiences an "epistemological barrier" between everyday and formal reasoning that often "keeps [her] from looking for connections between ideas from the different sides." | Jan chooses (consciously or unconsciously) (i) not to use knowledge in the classroom context that she did use in interview contexts and (ii) not to *reconcile* formal and informal knowledge in either context. | Role of epistemology in learning |
| Richards, "The Role of Affect in Sustaining Teachers' Attention and Responsiveness to Student Thinking" [45] | The affective experiences of Ms. L (fifth grade science teacher) and Ms. R (sixth grade science teacher) promoted and sustained their in-the-moment attention and responsiveness to student thinking. | Ms. L and Ms. R shifted their attention toward student thinking immediately following displays of curiosity/excitement and concern/frustration, respectively, and their responses to student thinking were plausibly linked to these affective experiences. | Role of teacher affect in teacher attention and responsiveness |

The claims illustrated by these four examples of case-oriented PER are interpretations of what is happening in a specific instance that draws on and has the potential to inform theory. Evidence for the claims are sequences of local events. Harlow's case shows the central role that adaptive instruction plays in elementary students' authentic scientific inquiry; Gupta, *et al.,* challenge the notion that misontologies are always unproductive for learning; Lising and Elby propose a mechanism that links epistemology and learning; and Richards bridges the literatures on affect and teacher attention/responsiveness to show that the former may play a role in stabilizing the latter. In each paper, the authors limit their claim(s) to the details of their individual cases; however, in providing a narrative and interpretations of their local events, these researchers show the audience what it can look like for a specific theory to "show up," in context, seeking to expand readers' vision for and understanding of their own contexts.

**Table II.** *Claims and evidence from examples of recurrence-oriented physics education research*

| Author and title of paper | Major claim(s) | Evidence for claim(s) | Significance: Informs predictions about? |
|---|---|---|---|
| Brewe, Traxler, de la Garza, and Kramer, "Extending positive CLASS results across multiple instructors and multiple classes of Modeling Instruction" [46] | "Modeling Instruction curriculum and pedagogy support the development of more favorable attitudes toward learning physics, independent of instructor." | Recurrence of attitudinal gains (as measured by the CLASS) across Modeling Instruction implementations | Factors that may contribute to positive attitudinal shifts |
| Hazari, Potvin, Lock, Lung, Sonnert, and Sadler, "Factors that affect the physical science career interest of female students: Testing five | "[D]iscussions about women's underrepresentation [in science] have a significant positive effect" on the physical science career interest of female | (1) Females who reported experiencing discussions of women's underrepresentation in their high school physics course were significantly more likely to express interest in pursuing a career in the physical sciences than were females who did not experience such discussions. (Both groups of females were | Classroom conditions that affect physical science career interest in females |
| common hypotheses" [47] | students; having "a single-sex physics class," "female physics teacher," and/or "female scientist guest speakers in science class" does not, nor does "discussing the work of female scientists in physics class." | statistically equivalent on the confounding variables.) (2) Females who reported (i) having a single-sex physics class, (ii) having a female physics teacher, (iii) having female scientist guest speakers in science class, or (iv) discussing the work of female scientists in physics class were statistically indistinguishable from females who did not, in terms of their expression of interest in pursuing a career in the physical sciences. (In each case, both groups of females were statistically equivalent on the confounding variables.) | |
| Koenig, Endorf, and Braun, "Effectiveness of different tutorial recitation teaching methods and its implications for TA training" [48] | *Tutorials* [8] recitation sections that use cooperative group work and Socratic instructor-student dialogue are more effective than those that use traditional lecture, individual group work, or cooperative group work with no Socratic dialogue. | (1) Introductory physics students who participated in (a) a recitation section that incorporated cooperative groups and Socratic dialogue between TAs and students performed significantly better on a conceptual post-test than did students who participated in recitation sections that incorporated (b) traditional lecture, (c) individual group *Tutorials* work, or (d) cooperative group *Tutorials* work (with no Socratic dialogue). (2) Fewer students experiencing style (a) "continued to use the same incorrect reasoning" on the post-test as on the pre-test, and more students in style (a) used the work-kinetic energy and impulse-momentum theorems on the post-test, compared to styles (b), (c), and (d). | Necessary components of physics teaching assistant training |
| Mikula and Heckler, "Student Difficulties With Trigonometric Vector Components Persist in Multiple Student Populations" [49] | (1) "[S]tudent difficulties with trigonometric vector components are persistent and pervasive…and [students' trigonometric vector component skills] are far below the requisite near-perfect accuracy needed for such fundamental skills." (2) Both (i) percentages of correct answers and (ii) percentages of certain errors in student reasoning depend on (a) the angle configuration and (b) the component of the vector requested. | (1) When asked to write an expression for the components of a vector after relevant course instruction, introductory physics students often made sign errors, interchanged sine and cosine, and "answer[ed] based on incorrectly drawn triangles and incorrectly placed angles." (2) Changing the angle configuration (while keeping the component requested the same) changes the (i) percentages of correct answers and (ii) percentages of certain errors in student reasoning. Changing the component requested (while keeping the angle configuration the same) changes the (i) percentages of correct answers and (ii) percentages of certain errors in student reasoning. | Difficulties students may have with vector components and factors that may affect difficulties |

In contrast to the claims listed in Table I, claims made by the examples of recurrence-oriented research in Table II are posed in terms of population-level patterns and relationships that have the potential to inform instructional predictions. Evidence for the claims includes the recurrence of research results and statistically significant differences between groups that are equivalent on all relevant measures but one. Brewe, *et al.,* demonstrate the effectiveness of a research-based curriculum in shifting students' attitudes about science; Hazari, *et al.,* determine what affects females' physical science career interest; Koenig, *et al.,* discern which of four recitation styles is most effective at improving student performance on written conceptual questions; and Mikula and Heckler report patterns and dependencies in student understanding of vector components. In drawing from random samples and/or privileging recurrence and reproducibility, these authors make claims about *populations,* seeking to inform predictions about the effectiveness of certain kinds of instruction for certain groups of students, or about what instructors may expect from these students.

## V. Theories of cause: On what grounds do researchers engaging in case-oriented PER make causal claims from single cases?

Our response to the question, "On what grounds do researchers engaging in case-oriented PER make causal claims from single cases?," is that case-oriented research assumes a theory of cause in which cause is inferred from process, rather than from controlled experiments. This perspective can be connected to the case-oriented research premises that action is shaped by the meaning-perspectives of participants and that causal mechanisms do not necessarily produce regularities. Meaning-making, and thus *cause* in the case-oriented research paradigm, is a process that is revealed in the details of specific cases as they unfold.

We see this question and its answer as particularly significant for the following reason: Even if a researcher who primarily identifies with recurrence-oriented PER accepts that universals can be discerned from the particulars of cases, he or she may still assume that causal claims can only be generated via controlled experimentation. However, this assumption is embedded within a theory of cause that is rarely shared by case-oriented research; thus, without understanding the distinct theory of cause associated with the case-oriented PER paradigm, these researchers may question the validity of – or at the least be confused by – the process by which causal claims are generated in case-oriented research.

In this section, we flesh out each of the assumptions that are tied to the choice to infer cause from process. We briefly address questions of the generalizability of causal claims inferred from single cases, tying this section back to the previous one, and we contrast examples of causal claims made by case- and recurrence-oriented PER, illustrating how the two theories of cause we describe are manifested in practice. We close by pointing to ways in which researchers often informally blend both theories of cause in their research, even if they consider one to be more rigorous or trustworthy.

### A. Social action is shaped by locally-constructed participant meanings

In case-oriented research, the *causes* of social action are participants' locally meaningful interpretations of their environments [13,50]. These meaning-perspectives are not only context-dependent but also dynamic, evolving as participants continually make sense of (and shape) their contexts and respond to other participants who are simultaneously making sense of (and shaping) the context [11,28,51]. Erickson [52] illustrates this with an analogy: "face-to-face social interaction is a process akin to that of climbing a tree that is climbing you back at the same time." Participant meanings often include non-physical entities, such as mental processes, that cannot be converted into variables without misrepresenting their true nature [53,54]. This perspective is echoed by the following quote:

> "When an historian asks 'Why did Brutus stab Caesar?' he means 'What did Brutus think, which made him decide to stab Caesar?' The cause of the event, for him, means the thought in the mind of the person by whose agency the event came about; and this is not something other than the event, it is the inside of the event itself."[6] [55]

This stance toward the social world was reflected in our interviews with physics education researchers. For example, one researcher described her ongoing analysis of interviews with middle school students:

---

[6] Case-oriented research that sought to understand why Brutus stabbed Caesar might connect this case to theory about, for example, some of the elements of violent political struggles or possible motives for murder.

> "There were many students who throughout the course of the interview would sometimes use chemistry vocabulary words. And then at other times, they would switch, where they would start drawing on their everyday experiences…And what I have been thinking about that is that that is an epistemological issue, so that actually, students aren't quite sure how to engage in these interviews that we use so regularly as researchers. And that they're sort of trying on a number of different ways that they could engage in the interview."

In this quote, the researcher attributes students' participation and knowledge-on-display to the meaning that they are making of their local context.

This differs from the perspective of recurrence-oriented research, in which the *causes* of human behavior are conceived in terms of (population-level) relationships between variables. This kind of research is modeled after that in the natural sciences, in which nature's "uniformity" allows a mechanical, chemical, or biological understanding of causation [13,20,27,28,34]. This "uniformity" does not imply linearity; rather, it implies that variation in human behavior follows a trend (*e.g.,* that the values for the parameters of interest are normally distributed across a population) [56,57]. Thus, any non-uniformities will likely average out to zero if one considers an entire population (or a large, representative sample of that population). This stance toward human behavior is represented in the following quote, in which a physics education researcher compares his research on teaching and learning to that in physics:

> "The way physics in general works is that you come up with a system of interest, you develop questions about that system, and then you develop ways to perturb the system and to measure the outcomes of the perturbation…And the general way we approach things, we know that systems are very complicated and that many things influence their dynamics, and we generally try to control the many things that influence the dynamics, understand them, to identify the factors, to control them, and then try to explore what varying one or two (a limited number) of those things, what effect that has. Where the ultimate objective is to understand the underlying dynamics…If you're successful in identifying the basic dynamics of the system, then you have the ability to hopefully influence it and affect it in ways that you desire to affect it. I'm trying to figure out the underlying dynamics [of physics learning and teaching], and yes, I think the basic approach is similar to the standard physics research approach, which is to try to understand the various factors involved in the system, to try to control a reasonable number of them and to vary certain others to look at the outcomes, with an aim to understanding the underlying dynamics."

This quote illustrates the sense that both the human and physical worlds can be modeled as governed by lawful relationships between variables, with the caveats that (1) there are many more variables to consider in social or cognitive interactions than in physical ones and (2) the interactions between these variables are much more complex.

### B. Mechanisms do not necessarily produce regularities

In the case-oriented paradigm, what is real to participants – and what shapes their actions – are the meanings they are making of their experiences. Because these meanings are inextricably linked to the local context, causal mechanisms do not necessarily produce regularities [11]. What is of interest in a causal account is how a *specific* event evolves and what mechanisms shape it [11,53] – those "processes that resulted in a specific outcome in a particular context" [58]. This is reflected by the quote in Section IV.C, in which an interviewee says that their work is about "mak[ing] a convincing case that that can be how things work, but not that it is how things would work next time."

Similarly, Berland and Hammer [10] caution against interpreting their results that certain framings are especially productive for argumentation as prescriptive. Instead, they suggest that instructors "design or build on situations that students find recognize as argumentative." In particular, they suggest making "finding something out" the starting place for instruction "and proceeding from there," implying that the specifics will emerge from students' meaning-making about a phenomenon, which will look different in different contexts.

In contrast, in the recurrence-oriented research paradigm, real phenomena – including relationships between variables – are reproducible. These relationships are thought to exist "independent of [scientists'] personal values and sociopolitical beliefs" [24] [see also [26]]. To ensure that one's observations and inferences *truly* reflect these relationships – and do not reflect biases that are the product of an anomalous, unrepresentative sample or of the personal values or beliefs of the researcher – researchers conducting recurrence-oriented PER privilege relationships that recur over and over, independent of observer and context [20]. Cook [21] ties *recurrence* to *truth*, saying that even though "observations are never theory-neutral, many of them have stubbornly re-occurred whatever the researcher's predilections" and thus have "such a high degree of facticity that they can be confidently treated as though they were true." What is of interest in a causal account are the context-independent relationships that can predict population-level behavior. The quote in Section IV.C reflects the sense that irregularities may reflect biases or random fluctuations; this interviewee says that he "begin[s] to believe [he's] onto something" when "[he] get[s] it a third time and…it's similar to what [he] observed the first two times."

Pollock and Finkelstein's [6] manuscript responds to this interviewee's call: one of the central questions of their study (and of the preceding manuscript on which it builds) concerns the reproducibility of large conceptual gains when *Tutorials* are implemented across and within institutions. The two graphs they display in their Figures 1 and 2 (see [6]) show the statistical indistinguishability of results from (1) (a) the University of Washington (*Tutorials* development site) and (b) CU-Boulder (*Tutorials* implementation site) *Tutorials* implementations and (2) the (i) first and (ii) second implementations of *Tutorials* at CU-Boulder. These graphs communicate that the gains achieved (1) at UW and (2) during the first implementation at CU-Boulder do not represent random fluctuations or irreproducible, extenuating circumstances; they represent a *real* curricular effect.

### C. Cause is inferred from a visualizable sequence of events that plausibly links local causes and effects

As illustrated by Figure 1, the case-oriented research perspective that social action is shaped by local participant meanings and that mechanisms do not necessarily produce regularities is connected to the choice to infer causality from "visualizable sequences of events, each event flowing into the next" [Maxwell [11] quoting Weiss [59]]. Smith [60] traces this to Gould [61], who writes:

> "…explanation takes the form of a narrative: E, the phenomenon to be explained, arose because D came before, preceded by C, B, and A. If any of those earlier stages had not occurred, or had transpired in a different way, then E would not exist (or would be present in a substantially altered form from E, requiring a different (but equally credible) explanation."

Thus, in this perspective, causality is "observed" when a sequence of events is connected by a plausible and compelling explanation (rooted in evidence from, for example, a video episode) for why one event follows the other. Because this theory deals with *local* causality – those events and processes that lead to specific outcomes – it lends itself well to case studies or other methods that use small numbers of individuals [11].

*Case-Oriented Research:*

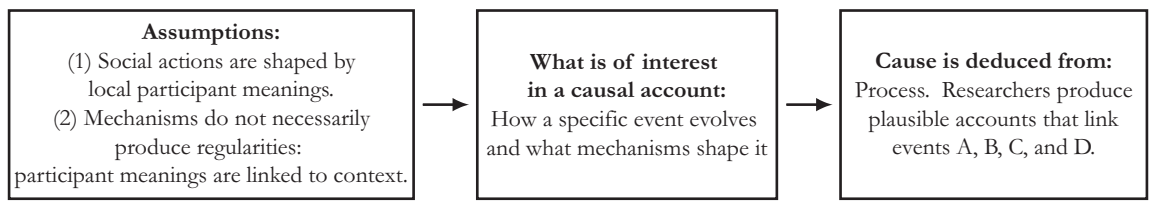

**Figure 1.** Simplified relationship between assumptions, focal points of causal accounts, and means by which cause is inferred in case-oriented research. The relationship between these three is more complex than this; each of the entities in the boxes may inform the other, and the arrows we have drawn can go the other way (*e.g.,* from what is of interest to the assumptions taken up).

For example, one researcher we interviewed described a causal claim that she inferred from process:

> "I think the thing I said was that Energy Theater *promotes,* and I think that…matters, because it's not a claim about frequency, it's a claim about process…I want to make a causal relationship between the Energy Theater representation and this disambiguating matter and energy. So I'm trying to say things like "because they were moving around material bodies in order to figure out what happened with the energy in the light bulb…see how…the ropes…assisted them in thinking about such-and-such. And had it been a bar chart, like, see how that really could only have happened because of the way Energy Theater is, right." So…because my claim is a *promotes,*…if I can show like a plausible causal relationship, even in one good case, if I can do it well enough that you…the reader are like, "Wow, yeah, it is the ropes, isn't it?," then I don't have to show you ten examples of it being the ropes, because you get it, you believe that it's because of the ropes and you see easily that it would be because of the ropes."

When this researcher says "see how the ropes assisted them in thinking about such-and-such," she refers to a sequence of events in which teachers' interactions with ropes (a part of the Energy Theater representation) play a role in the disambiguation of matter and energy.

Berland and Hammer [10] make several causal claims, all deduced from process. In particular, they explain the dynamics of each episode in terms of framings – which embed particular social and epistemic expectations – that are co-constructed and dynamically negotiated by participants. For example, the authors attribute the discord in the third episode to inconsistent expectations among students and teacher. Figure 2 outlines their argument: In the midst of the argumentative discussion, in which students stably assume social and epistemic authority (and the teacher reinforced this framing), Mr. S bids to "move on" and walks to the front of the room, signaling a framing shift and a bid for others to recognize *him* as social and epistemic authority. Several students take up his bid, requesting answers from Mr. S and cheering when he offered them. Others, however, maintain the argumentative discussion frame, sustaining their social and epistemic authority by talking out of turn and challenging or questioning Mr. S' claims, which "le[a]d[s] to instabilities". As they do so, Mr. S dismisses their challenges, pressing them to accept his answer, and disciplines students who talk out of turn, reinforcing his framing of himself as social and epistemic authority. The authors summarize by suggesting "that the tension that emerged in the discordant discussion *resulted from* the combination of more traditional school framings and those that align with scientific argumentation" (emphasis added).

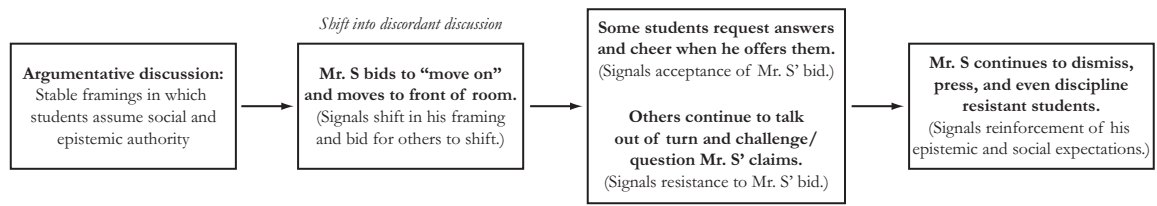

**Figure 2.** Outline of Berland and Hammer's [10] causal claim, which connects actions to evolving framings

In short, the authors attribute the discord that originally captured their attention to the *meanings* that participants make of their participation and authority in this discussion. These meanings evolve dynamically, shifting or stabilizing (or even strengthening) in response to verbal and nonverbal signals communicated between participants (*e.g.,* Mr. S' dismissal and discipline of particular students was a response to their resistance of his original bid, and they were perceived as resistant in the context of Mr. S' shifting his framing of the discussion).

In contrast, the recurrence-oriented research perspective that human behavior is guided by relationships between variables and that mechanisms do produce regularities is connected to the choice to infer cause from controlled experiments, as in Figure 3.

*Recurrence-Oriented Research:*

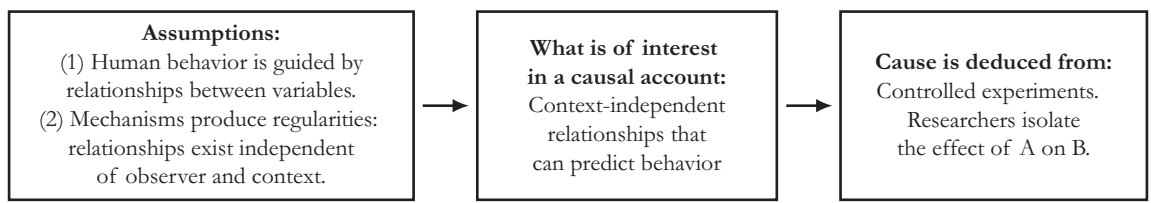

**Figure 3.** Simplified relationship between assumptions, focal points of causal accounts, and means by which cause is inferred in recurrence-oriented research. As in Figure 1, the relationship between these three is more complex than this; each of the entities in the boxes may inform the other, and the arrows we have drawn can go the other way (*e.g.,* from what is of interest to the assumptions taken up).

Inferring population-level causal relationships not only requires random sampling (to ensure that the sample is representative of the population at large); it also requires that treatment and control groups are statistically equivalent on all measures but the variable being tested (or that the effect of confounding variables is measurable via some other means) [32,57]. The latter ensures that the difference(s) between the treatment and control group can be attributed to the treatment, and not to some other variable that has not been taken into account [62]. Researchers use experimental techniques – such as random assignment to treatment and control groups – and statistical methods – such as multi-variate regression analysis – to meet this requirement [21,37,56].

This orientation toward cause as inferred from controlled experiments was reflected in our interviews with physics education researchers and in our example of published recurrence-oriented PER. For example, one researcher described his work as

> "trying to study in detail why students are answering the way that they do. I'm very agnostic about that, and I try to take an extremely empirical approach…I try…as much as possible…to collect data…that people would regard as reliable and reproducible."

He goes on to say that typically, his research involves both reporting patterns in student responses to questions about topic *X* and trying to understand the variables that might affect students' responses to these questions. He pursues the latter via controlled experiments. For example, to understand which variables affect student responses to questions about two-dimensional 'time of flight', this researcher and his colleagues first devise questions that illustrate trajectories with different characteristics (*e.g.,* different heights, ranges, areas under the curve). (In other words, they choose relevant variables.) They ask each question in multiple contexts to ensure that students interpret the question the way it was meant to be interpreted, eliminating questions that are idiosyncratic. Drawing from the remaining questions, they show students two trajectories at a time, each with different characteristics, and ask them to choose the one that has the longest time of flight. The characteristics (inputs) are varied independently and simultaneously, and researchers watch to see how the patterns in student responses (output) change. He describes the results as follows:

> "Students tend to say that if something goes really, really far, it takes really, really long, independent of whether it goes high or not…[T]hose things come into play when they're making a decision, because you can see that when you change these things, their answer changes."

Pollock and Finkelstein's [6] manuscript serves as an example of published research that embodies this approach to causal questions and processes, using a *post hoc* experimental design. The authors make two (primary) causal claims: (1) that student learning gains are impacted by the curriculum, and (2) that student learning gains rely on faculty background. These claims are deduced from the covariation of the normalized gain <g> on a standardized assessment with the recitation curriculum ("all courses with tutorial experiences lead to learning gains higher than all classes that have traditional recitation sections"), and with faculty background (there is a "sizable variation of success among these implementations"). They posit reasons for the "potential effect of faculty", such as that PER faculty may be more familiar with the development of the reformed curriculum and may therefore implement it in a way that is better aligned with curricular goals. Pollock and Finkelstein's manuscript further instantiates this theory of cause by claiming that (emphasis added) "the *most compelling [evidence* that curriculum choice matters] would be to *control for faculty member and vary the curricula.* However, we have not had opportunity to [do so]." They go on to cite the recurrence of positive gains across instructors as evidence for this relationship in the absence of controlled experiments.

### D. Generalizations are made by connecting local mechanisms to theory.

Researchers engaging in case-oriented PER infer causal claims from visualizable sequences of events, assuming a theory of cause in which social actions shape local participant meanings and in which mechanisms do not necessarily produce regularities. Researchers make *generalizable* claims from single cases by connecting local mechanisms to theory [5,11,33]. Maxwell [11] states that research efforts are "most productive if they are informed by, and contribute to, a detailed theory…of the causal process being investigated." For example, Berland and Hammer's [10] interpretation of the sequence of events in Mr. S' classroom is informed by theory on framing and theory on argumentation, and the relationship between these theories that emerges from their analysis – that certain framings may be more productive for argumentation – contributes to both theoretical spaces.

This contrasts with the practices of recurrence-oriented PER, in which researchers generalize causal relationships by replicating experiments or by observing the recurrence of causal relationships across contexts [21,62,63]. For example, Pollock and Finkelstein's manuscript [6] reproduces a curricular effect, showing that the *Tutorials* are generally effective (*i.e.,* beyond non-standard educational contexts).

### E. Process-oriented theory of cause addresses different research questions than experimentally-oriented theory of cause

Just as answering to different perspectives on generalization means that case- and recurrence-oriented research generate different kinds of claims, answering to different theories of cause means that these two paradigms generate different kinds of *causal* claims. Researchers who question the possibility of making causal claims on the basis of a single case may expect all research to establish population-level relationships that are predictive of other events. In this section, we revisit the additional examples of published case- and recurrence-oriented PER introduced in Section IV.E, highlighting those that make causal claims and indicating the means by which these claims were inferred. Our goal is to provide examples of the kinds of causal claims that case-oriented research can generate, to show how these are tied to the case-oriented theory of cause we have articulated in this section, and to contrast these claims with examples that are associated with the recurrence-oriented theory of cause.

As we discussed in Section IV.E, the causal claims made by the three manuscripts in Table III are about what is happening in a specific instance that draws on and has the potential to inform theory. In each case, the claims propose a mechanistic relationship that accounts for a series of events in a classroom context. For example, Gupta, Elby, and Conlin examine the evolution of one group of teachers' discourse to infer the productive role that a "misontology" plays in their understanding of gravity; Lising and Elby infer a causal relationship from the way in which Jan engages in a series of events, both in the classroom and in interviews; and Richards analyzes sequences of classroom events to show how specific affective experiences are initiated and then sustain teacher attention to student thinking.

**Table III.** *Causal claims made by additional examples of case-oriented PER*

| Author and title of paper | Causal claim(s) | Cause was inferred from: |
|---|---|---|
| Gupta, Elby, and Conlin, "How substance-based ontologies for gravity can be productive: A case study" [43] | "…[Lynn's] Galilean reasoning emerged *because of,* not in spite of, the teachers' misontologies of gravity. And this idea of each coin in a roll feeling a certain amount of gravity then *fed into* the Newtonian compensation argument whereby the heavier object feels more gravitational pull but also puts up more resistance to getting moved." (emphases added) | Evolution of classroom discourse |
| Lising and Elby, "The impact of epistemology on learning: A case study from introductory physics" [44] | Jan's experience of a "epistemological barrier" between everyday and formal reasoning that often "*keeps [her]* from looking for connections between ideas from the different sides." (emphasis added) | Series of classroom and interview events, refutation of alternative explanations |
| Richards, "The Role of Affect in Sustaining Teachers' Attention and Responsiveness to Student Thinking" [45] | The affective experiences of Ms. L and Ms. R *stabilize* their attention and responsiveness to student thinking. | Sequence of classroom events |

In contrast, the claims made by the recurrence-oriented papers listed in Table IV are not accounts of what is happening in a specific instance but instead are reports of relationships between two variables for a particular population (*e.g.,* female college students or introductory physics students). Koenig, *et al.,* and Mikula and Heckler each plan and conduct controlled experiments to test relationships between variables; Hazari, *et al.,* use multi-variate matching techniques [64,65] to compose control and treatment groups after the fact; and Brewe, *et al.,* use the consistency of specific outcomes to propose plausible mechanisms for positive attitudinal shifts. Brewe, *et al.,* are careful to qualify that the co-occurrence of certain plausible causes and effects is "not adequate to draw causal conclusions regarding to what specifically the shifts should be attributed" but can "provide insight into the factors that could mechanistically explain" the shifts.

**Table IV.** *Causal claims made by additional examples of recurrence-oriented PER*

| Author and title of paper | Causal claim(s) | Cause was inferred from: |
|---|---|---|
| Brewe, Traxler, de la Garza, and Kramer, "Extending positive CLASS results across multiple instructors and multiple classes of Modeling Instruction" [46] | Students' positive attitudinal shifts:** (1) Can be "attribute[d]" to Modeling Instruction. (2) Likely do not "ar[i]se from a 'good semester' or any unique expertise of the professor." (3) May be attributable to small class sizes. (4) May be promoted by explicit focus on epistemological resources. | Consistent co-occurrence of variables of interest (or lack thereof) |
| Hazari, Potvin, Lock, Lung, Sonnert, and Sadler, "Factors that affect the physical science career interest of female students: Testing five common hypotheses" [47] | "[D]iscussions about women's underrepresentation [in science] *have a significant positive effect*" on the physical science career interest of female students; having "a single-sex physics class," "female physics teacher," and/or "female scientist guest speakers in science class" does not, nor does "discussing the work of female scientists in physics class." (emphasis added) | Multi-variate matching methods that isolates variables of interest |
| Koenig, Endorf, and Braun, "Effectiveness of different tutorial recitation teaching methods and its implications for TA training" [48] | The "manner in which the *Tutorials* [8] are taught" *affects* student understanding. | Controlled experiment that isolates variables |
| Mikula and Heckler, "Student Difficulties With Trigonometric Vector Components Persist in | (i) Percentages of correct answers and (ii) percentages of certain errors in student reasoning *depend on* (a) the angle configuration and (b) the component of the vector requested. | Controlled experiment that isolates variables |

| Multiple Student Populations" [49] | | |

When comparing the claims made by case- and recurrence-oriented research, one point of confusion may be that each case-oriented claim can be rephrased in terms of a relationship between variables: ontologies affect learning, epistemology affects learning, and affect affects attention. Readers may wonder how these claims are different than those made by recurrence-oriented research. This brings up an important point: both case- and recurrence-oriented PER are seeking mechanisms that explain teaching and learning phenomena, and the product (claims) of both is often a mechanism. What differs between the two is where researchers expect mechanisms to appear (at the level of the population versus the case), the ways in which they are expected to generalize (to a population versus to theory), and how they expect to use them (to make predictions versus to broaden readers' awareness). When a researcher conducting case-oriented PER reports that "epistemology affects learning," they have likely brought the lens of epistemology to bear on an instance of learning and seen that it can explain how participants are making meaning of their experiences. They expect this lens to be useful in other contexts, but in ways that are intimately tied to these other contexts. The mechanism is not expected to produce regularities. On the other hand, when a researcher conducting recurrence-oriented research reports that "a particular intervention shifts students' attitudes about science," they have likely shown that this mechanism explains regularities in their data, and they expect it to continue to do so, such that readers can predict the effectiveness of the intervention for other students who are members of the same population.

### F. Informal blending of theories of cause

We suspect that many physics education researchers informally blend elements of both theories of cause in their research. For example, when Robertson conducted recurrence-oriented PER that emphasized student conceptual understanding and curriculum development, she often situated herself as an instructor or observer in courses that used her curriculum in order to watch the implementation process unfold. These observations of process would then inform changes to the curriculum or new research directions, suggesting that she (at least implicitly) understood the way that those events unfolded as significant in answering the question of *why* the curriculum was or was not effective. However, she typically did not consider this a part of her formal research, referring to it in writing as "anecdotal" evidence [66]. Researchers engaged in case-oriented physics education research, meanwhile, may describe their research as "discovering relevant variables." This suggests to us that these researchers see relationships between variables as significant in understanding why an event happens the way that it does, even if they do not see determination of those relationships as part of their own research. To the extent that individual researchers more strongly identify with one theory of cause or the other, they may tend to see the questions inspired by that theory as more rigorous, trustworthy answers to the question of *why*.

## VI. Discussion and conclusions

A primary goal of this research is to translate between paradigms in order to foster conversation and understanding among researchers who may take fundamentally different perspectives toward research. This paper aims to make visible some of the critical, relevant questions and practices of case-oriented physics education research, in particular, and to substantiate its methodological rigor alongside that of recurrence-oriented PER.

We situate this manuscript in the case-oriented research paradigm, since much of our work has comprised connecting the practices and perspectives articulated by active members of our field to one another and to the theory and practices of qualitative and quantitative social sciences research more generally. We hope that our work will be useful to members of our field in making sense of and engaging with research outside the paradigm with which they most strongly resonate. In the tradition of case-oriented research, we expect the specifics of these themes of selection, generalization, and theories of cause to vary by context while at the same time pointing back to some of the fundamentals we have described here.

Understanding the case-oriented and recurrence-oriented PER paradigms – and the meanings that researchers make as they instantiate these paradigms – has enhanced our own vision and broadened our awareness as we engage in conversations about PER. It has prompted us, for example, to seek to connect the theoretical perspectives articulated in examples of case-oriented PER to the specific cases these examples describe. It has given us the freedom to imagine possible applications of this theory to our own contexts without expecting identical results. It has supported us in valuing the priority of recurrence in examples of recurrence-oriented PER and in using these research results to make reasonable predictions about what to expect in our own classrooms. In short, understanding the rationale and practices of selection, generalization, and causal-claims-generation in case-oriented and recurrence-oriented physics education research has given us the tools to evaluate, appreciate, and better understand research from outside the paradigm with which we most strongly identify.

We suggest that our analysis is particularly significant for practices of peer evaluation in PER. If case-oriented and recurrence-oriented research make different assumptions about the social world and research rigor, then these *different research paradigms call for different evaluative criteria.*[7] In evaluating research across paradigms, we recommend that evaluators strive to first understand the authors' selection criteria, approach to generalization, and theory of cause, recognizing that they may be different than their own. The following questions may be helpful in doing so:

1. *Is the researcher's selection processes made clear?* Do the researchers articulate their aims, theoretical commitments, interests, and beliefs about knowledge and social phenomena? Do they clarify why the relevant selections were made, and were these selections reasonable, given the researcher's commitments? Can the reader recognize the interpretive categories, variables, codes, or the theoretical importance of the selected data once the authors have described them?

2. *In what sense is the work generalizable?* If the research is case-oriented, has the author connected the case to a broader theory? Does the research have the potential to broaden awareness of similar cases? If the research is recurrence-oriented, has the author demonstrated representativeness or reproducibility across contexts? Are the parameters and relationships described relevant to many contexts?

3. *To which implicit theory of cause does the research subscribe, and are the claims and evidence consistent with this theory?* In other words, if a causal claim is made, does the researcher treat cause as meanings made by participants or as a relationship between variables? Are the claims about cause connected to visualizable sequences of events or to the results of controlled experiments? In general, does the evidence, paired with the implicit theory of cause, support the claims that were made?

We suggest that researchers reporting case-oriented PER support the broader PER community in appreciating, understanding, and evaluating their research by explicitly articulating: (1) their selection practices, including the influence of their research aims and assumptions; (2) the means by which they generalize their results and for what purposes they expect their work to be useful; and (3) the theory of cause implicit in any causal claims they make.

---

[7] This perspective is echoed by a recent call from the American Educational Research Association [70] (cited by [16]) that the methodological adequacy of a manuscript be judged "in relation to the 'methodological requirements of its type…[and] the significance of its result in the context of the problems internal to its own tradition, and not the requirements and aspirations of types to which it does not belong."

# VII. Acknowledgments

This material is based upon work supported by the National Science Foundation under Grant Nos. 0822342 and 122732. We are deeply indebted to each of the physics education researchers that shared their thoughts with the first author and for feedback on this manuscript from James Day, Paula R. L. Heron, David E. Meltzer, Justin C. Robertson, Eleanor Sayre, Thomas M. Scaife, and Michael C. Wittmann. We also appreciate the thoughtful feedback from members of the Energy Project team, particularly Stamatis Vokos and from three anonymous reviewers, as well as from members of the Physical Review Special Topics – Physics Education Research Editorial Board.